\documentclass[conference]{IEEEtran}
\IEEEoverridecommandlockouts
\usepackage{cite}
\usepackage{amsmath,amssymb,amsfonts}
\usepackage{algorithmic}
\usepackage{graphicx}
\usepackage{mathtools}
\usepackage{booktabs}
\usepackage{textcomp}
\usepackage{xcolor}
\usepackage{hyperref}
\usepackage{soul}
\usepackage{tikz}
\usetikzlibrary{positioning, arrows.meta, fit, calc, shapes.geometric, backgrounds}

\newcommand{\diff}{\mathrm{d}}

\def\BibTeX{{\rm B\kern-.05em{\sc i\kern-.025em b}\kern-.08em
    T\kern-.1667em\lower.7ex\hbox{E}\kern-.125emX}}
\begin{document}

\title{A Variational Lagrangian Framework for \\ Log-Homotopy Particle Flow Filters\\
{}
\thanks{This work was funded by the National Research, Development and Innovation Office (NKFIH) under NKKP Grant Agreement No. ADVANCED 152880.}
}

\makeatletter
\newcommand{\linebreakand}{%
  \end{@IEEEauthorhalign}
  \hfill\mbox{}\par
  \mbox{}\hfill\begin{@IEEEauthorhalign}
}
\makeatother
\author{
\IEEEauthorblockN{
1\textsuperscript{st} Olivér Törő\IEEEauthorrefmark{1}, 
2\textsuperscript{nd} Domonkos Csuzdi\IEEEauthorrefmark{2}, and 
3\textsuperscript{rd} Tamás Bécsi\IEEEauthorrefmark{3}
}
\IEEEauthorblockA{\textit{Department of Control for Transportation and Vehicle Systems} \\
\textit{Faculty of Transportation Engineering and Vehicle Engineering}\\
\textit{Budapest University of Technology and Economics} \\
Muegyetem rkp. 3., Budapest, 1111, Hungary}
\IEEEauthorblockA{\IEEEauthorrefmark{1}Email: toro.oliver@kjk.bme.hu, ORCID: 0000-0002-7288-5229, corresponding author}
\IEEEauthorblockA{\IEEEauthorrefmark{2}Email: domonkos.csuzdi@edu.bme.hu, ORCID: 0000-0003-4774-3330}
\IEEEauthorblockA{\IEEEauthorrefmark{3}Email: becsi.tamas@kjk.bme.hu, ORCID: 0000-0002-1487-9672}
}

\maketitle

\begin{abstract}
The log-homotopy particle flow filter resolves the Bayesian update by transporting particles along a continuous trajectory in pseudo-time. However, the governing partial differential equation for the flow velocity is fundamentally underdetermined, admitting an infinite family of valid solutions. In this work, we regard the particle flow as the motion of a pressureless inviscid fluid. We define a Lagrangian action based on the kinetic energy of the system, subject to the constraints imposed by the continuity equation and the log-homotopy evolution. By applying the principle of least action, we obtain the Euler--Lagrange equations for the optimal flow, which yields an irrotational potential flow structure. We show that this variational framework yields a coupled Hamilton--Jacobi equation structurally isomorphic to Madelung's hydrodynamic formulation of quantum mechanics. In this analogy, the log-homotopy constraint acts as a generalized quantum potential that generates the force required to guide the probability fluid along the exact Bayesian update path. Finally, we derive the material acceleration of the flow, shifting the formulation from a kinematic to a dynamical description. This perspective could enable the application of higher-order symplectic integrators for improved numerical stability and provide a physics-based metric for adaptive stiffness detection in high-dimensional filtering.
\end{abstract}

\begin{IEEEkeywords}
particle flow, Bayesian update, log-homotopy, Lagrangian
\end{IEEEkeywords}

\section{Introduction}

In the framework of Bayesian inference, the recursive estimation of the state of a nonlinear dynamic system is challenged by the intractability of high-dimensional integration. While sequential Monte Carlo (SMC) methods \cite{chen2012monte,robert2004monte}, commonly known as particle filters \cite{gordon1993novel}, provide a flexible non-parametric approximation of the posterior density, their practical application in high-dimensional state spaces is affected by the phenomena known as weight degeneracy and the curse of dimensionality \cite{daum2003curse,bengtsson2008curse}. As the state space dimension increases, the volume of the state space expands exponentially, causing the
likelihood function to become extremely peaked relative to the
prior. Consequently, the vast majority of particles are assigned negligible weights, and the effective sample size collapses to zero. This leads to frequent resampling that depletes statistical diversity \cite{snyder2008obstacles}.

To mitigate the issues of particle depletion and weight collapse caused by the assimilation of highly informative observations, various progressive Bayesian update methods have been proposed. The underlying principle of these approaches is to avoid the pointwise multiplication of the prior by the full likelihood function in a single discrete step. Instead, the measurement is assimilated through a sequence of numerically stable partial updates. An early realization of this concept was introduced by Oudjane and Musso \cite{oudjane2000progressive}, who proposed a progressive correction for regularized particle filters by factorizing the likelihood function into a sequence of intermediate, less peaked sub-likelihoods. Similarly, methods such as annealed importance sampling \cite{neal2001annealed} and SMC samplers \cite{del2006sequential} construct a geometric path of intermediate distributions, typically bridging the prior and posterior via an inverse temperature parameter.

Hanebeck et al. \cite{hanebeck2003progressive} introduced the progressive Bayes framework, where the probability density evolves according to a system of linear first-order ordinary differential equations defined over an artificial progression parameter. This formulation minimizes the squared integral deviation between the true and approximated density. Subsequent works successfully extended this homotopy-based approach to polynomial nonlinearities and moment calculations \cite{huber2013gaussian}.

Daum and Huang introduced the log-homotopy particle flow filter approach in 
\cite{daum2007nonlinear}. Unlike standard importance sampling, this method transports particles along a continuous trajectory in pseudo-time ($\lambda$), governed by a stochastic differential equation. The flow is constructed such that the particle distribution exactly tracks a log-homotopy connecting the prior to the posterior. However, the governing partial differential equation (PDE) for the flow velocity is fundamentally underdetermined. The continuity equation, which ensures mass conservation, provides only a single scalar constraint for the multi-dimensional velocity field, admitting an infinite family of valid solutions.

Existing literature has addressed this non-uniqueness through various heuristic constraints. The earliest log-homotopy particle flow variant, the incompressible flow \cite{daum2007nonlinear, Daum2008particle}, assumed that the total derivative of log-density with respect to $\lambda$ is zero, yielding an incompressible flow. This was followed by the exact flow variant \cite{daum2010exact,ding2012implementation}, which yields a closed-form solution assuming a Gaussian prior and a linear measurement model with additive Gaussian noise. Subsequent variants introduced non-zero diffusion terms to improve numerical stability \cite{Daum2013Particle,Daum2016Seven,daum2018new}, or explored the possibility of using deep learning architectures to approximate the flow \cite{csuzdi2026physics} based on the principles of physics-informed neural networks (PINN) \cite{RAISSI2019686}.
However, whether tackled via heuristic analytical constraints or black-box optimization frameworks, the fundamental physical and geometric nature of the optimal transport map in these particle flows remains underexplored.

In this work, we frame the particle transport within a variational Lagrangian framework. Drawing inspiration from the fluid dynamic formulation of optimal transport \cite{benamou2000computational}, we model the probability mass as a pressureless inviscid fluid. We define an action functional corresponding to the total kinetic energy of the system and seek the velocity field that minimizes this action, subject to the exact physical constraints of the Bayesian update. 

In this paper, we formulate the exact particle flow as a constrained variational problem and derive the associated Euler--Lagrange equations. The condition of minimal kinetic energy mathematically ensures an irrotational potential velocity field.
An isomorphism is shown between the potential log-homotopy particle flow and Bohmian mechanics (Madelung hydrodynamics). We demonstrate that the Bayesian update induces an information-theoretic guiding potential that is structurally identical to the quantum potential. It exerts a force required to steer the macroscopic probability fluid. We derive the explicit material acceleration (Euler equation) of the particle system from the coupled Hamilton--Jacobi formulation. Importantly, the primary motivation for introducing the variational formulation is not merely energetic optimality, but the induced Hamiltonian structure itself. By expressing the particle flow in canonical form, the framework provides access to analytical tools from classical mechanics, including Hamilton--Jacobi theory and canonical transformations. It also enables the use of structure-preserving symplectic integration methods for exact or numerically stable flow propagation. This shifts the focus from purely kinematic mappings to a dynamical perspective, providing a theoretical basis for higher-order symplectic integrators and adaptive stiffness detection in high-dimensional particle flows.

The structure of the paper is the following. Section~\ref{sec:problem} presents the problem formulation of the log-homotopy particle flow method. The variational Lagrangian approach is presented in Section~\ref{sec:var}. Theoretical analysis of the flow equations and potentials is discussed in Section~\ref{sec:interpret}. Section~\ref{sec:exp} presents numerical examples to demonstrate the effect of generating the flow from a potential. Conclusion and future work are presented in Section~\ref{sec:conc}.

\section{Problem formulation} \label{sec:problem}

\subsection{Preliminaries and Notation}
The state space is the $d$-dimensional Euclidean space $\mathbb{R}^d$. For a scalar function $f(x) : \mathbb{R}^d \rightarrow \mathbb{R}$, its gradient is denoted $\nabla f$. For a vector field $v(x) : \mathbb{R}^d \rightarrow \mathbb{R}^d$, its divergence is denoted $\nabla \cdot v$.

We focus on the one-step Bayesian inference problem. Let $x \in \mathbb{R}^{d}$ denote the random state vector and $z \in \mathbb{R}^{m}$ the measurement vector. Let $g(x)$ denote the prior probability density function (pdf) and $h(x) \triangleq p(z|x)$ the likelihood of the measurement $z$. By Bayes' theorem, the posterior pdf $p(x)$ is given by:
\begin{equation}
    p(x) = \frac{g(x)h(x)}{K} \, ,
\end{equation}
where $K = \int g(x)h(x) \diff x$ is the normalization constant. We assume all pdfs are sufficiently smooth and nowhere-vanishing.

\subsection{The Log-Homotopy Evolution}
Particle flow filters resolve the Bayesian update by transporting samples from $g(x)$ to $p(x)$ continuously over a pseudo-time interval $\lambda \in [0, 1]$. A path of intermediate densities $p(x, \lambda)$ is constructed using a linear log-homotopy:
\begin{equation} \label{eq:homotopy}
    \log p(x, \lambda) = \log g(x) + \lambda \log h(x) - \log K(\lambda) \, ,
\end{equation}
where $K(\lambda) = \int g(x) h(x)^\lambda \diff x$ is the instantaneous normalization factor. 

\subsection{The Continuity Constraint}
We model the probability transport as the flow of a pressureless fluid. The evolution of the density $p(x, \lambda)$ and the flow velocity $v(x, \lambda)$ must satisfy the continuity equation:
\begin{equation} \label{eq:continuity}
    \frac{\partial p}{\partial \lambda} + \nabla \cdot (p v) = 0 \, .
\end{equation}
Substituting \eqref{eq:homotopy} into \eqref{eq:continuity} yields the fundamental constraint for the exact flow \cite{daum2011coulomb,csuzdi2026physics}:
\begin{equation} \label{eq:master_eqn}
    -\nabla \cdot (p v) = p \left( \log h(x) - \mathbb{E}_{p}[\log h(x)] \right) \, .
\end{equation}
Equation \eqref{eq:master_eqn} is a linear partial differential equation for $v$. However, it is underdetermined for dimensions $d > 1$ since it provides a single scalar constraint for the $d$ components of $v$. Consequently, there exist infinitely many valid velocity fields that generate the same marginal density evolution. The variational formulation introduced in the next section can be interpreted as resolving this non-uniqueness by selecting a distinguished irrotational representative from the equivalence class of admissible flow fields through a minimum-action principle.

The transport of the particles is governed by an ordinary differential equation (ODE):
\begin{equation}\label{eq:ode}
\frac{\diff x}{\diff \lambda} = v(x, \lambda). 
\end{equation}
While analytic solutions exist for specific, restricted cases \cite{toro2023analytic}, particle trajectories are typically obtained via numerical integration. Because the governing ODE \eqref{eq:ode} in log-homotopy filters often suffers from stiffness \cite{crouse2021particle}, maintaining filter performance requires specialized techniques, such as adaptive Euler step-size selection \cite{mori2016adaptive} or the incorporation of a regularizing scheme \cite{fok2023mpi}.

\section{Variational derivation of the optimal flow} \label{sec:var}

To resolve the ambiguity of \eqref{eq:master_eqn}, we apply the principle of least action. We seek the specific velocity field that minimizes the kinetic cost of transport while strictly adhering to the Bayesian update.

\subsection{Lagrangian action functional}
We define the Lagrangian density $\mathcal{L}_0$ as the kinetic energy of the probability fluid:
\begin{equation}
    \mathcal{L}_0(x, \lambda) = \frac{1}{2} p(x, \lambda) \|v(x, \lambda)\|^2 \, .
\end{equation}
The optimization task is to minimize the total action $S$ subject to two constraints: mass conservation and the log-homotopy condition. Two scalar Lagrange multiplier fields, $\phi(x, \lambda)$ and $\psi(x, \lambda)$, enforce these constraints respectively.

The constrained action functional is:
\begin{equation} \label{eq:total_action}
    S[p, v, \phi, \psi] = \int_0^1 \int_{\mathbb{R}^{d}} \Big( \mathcal{L}_0 + \phi c_1 + \psi c_2 \Big) \diff x \diff \lambda \, ,
\end{equation}
where the constraint terms are defined as:
\begin{align}
    c_1(x, \lambda) &= \frac{\partial p}{\partial \lambda} + \nabla \cdot (p v) = 0 \, , \\
    c_2(x, \lambda) &= \log p - \log g - \lambda \log h + \log K = 0 \, .
\end{align}
We assume that $p$ and $v$ decay sufficiently fast as $\|x\| \to \infty$ to ensure surface integrals at the boundary vanish.

\subsection{First Variation: Optimal Velocity Structure}
We first compute the variation of $S$ with respect to the velocity $v$. Isolating terms dependent on $v$ in \eqref{eq:total_action}:
\begin{equation}
    S_v = \int_0^1 \int_{\mathbb{R}^{d}} \left( \frac{1}{2}p \|v\|^2 + \phi \nabla \cdot (p v) \right) \diff x \diff \lambda \, .
\end{equation}
We take the variation $v \to v + \delta v$:
\begin{equation}
    \delta S_v = \int_0^1 \int_{\mathbb{R}^{d}} \left( p v \cdot \delta v + \phi \nabla \cdot (p \delta v) \right) \diff x \diff \lambda \, .
\end{equation}
We apply the identity $\phi \nabla \cdot A = \nabla \cdot (\phi A) - A \cdot \nabla \phi$ to the second term, setting $A = p \delta v$. By the divergence theorem, $\int \nabla \cdot (\phi p \delta v) \diff x$ becomes a boundary integral, which vanishes under our decay assumptions. The variation simplifies to:
\begin{equation}
    \delta S_v = \int_0^1 \int_{\mathbb{R}^{d}} p (v - \nabla \phi) \cdot \delta v \diff x \diff \lambda = 0 \, .
\end{equation}
Since this equation must hold for arbitrary perturbations $\delta v$, and assuming $p > 0$, we obtain:
\begin{equation} \label{eq:potential_flow}
    v(x, \lambda) = \nabla \phi(x, \lambda) \, .
\end{equation}
This result shows that the minimum-action admissible flow is necessarily irrotational and admits a scalar potential representation. The velocity is fully determined by the gradient of the scalar potential $\phi$.

\subsection{Second variation: Hamilton--Jacobi dynamics}
Next, we perform the variation with respect to the density $p$. The relevant terms in the action are:
\begin{equation}
    S_p =\! \int \!\! \int \left[ \frac{1}{2} \|v\|^2 p + \phi \left( \frac{\partial p}{\partial \lambda} + \nabla \cdot (p v) \right) + \psi \log p \right]\! \diff x \,\diff \lambda .
\end{equation}
Setting the variation $\delta S_p = 0$:
\begin{equation}
    \int \!\! \int \left[ \frac{1}{2} \|v\|^2 \delta p + \phi \frac{\partial \delta p}{\partial \lambda} + \phi \nabla \cdot (v \delta p) + \frac{\psi}{p} \delta p \right] \diff x \diff \lambda = 0.
\end{equation}
We integrate the time-derivative term by parts with respect to $\lambda$ and the divergence term by parts with respect to $x$. Assuming fixed initial/terminal conditions ($\delta p|_{\lambda=0,1}=0$) and vanishing boundary terms at infinity:
\begin{align}
    \int_0^1 \phi \frac{\partial \delta p}{\partial \lambda} \diff \lambda &= - \int_0^1 \frac{\partial \phi}{\partial \lambda} \delta p \diff \lambda \, , \\
    \int_{\mathbb{R}^{d}} \phi \nabla \cdot (v \delta p) \diff x &= - \int_{\mathbb{R}^{d}} (v \cdot \nabla \phi) \delta p \diff x \, .
\end{align}
Substituting these results back yields:
\begin{equation}
    \int_0^1 \!\int_{\mathbb{R}^{d}} \!\left( \frac{1}{2} \|v\|^2 - \frac{\partial \phi}{\partial \lambda} - v \cdot \nabla \phi + \frac{\psi}{p} \right) \delta p \,\diff x \,\diff \lambda = 0 \, .
\end{equation}
For this to hold for all admissible $\delta p$, the integrand must vanish. Using $v = \nabla \phi$, we obtain the Hamilton-Jacobi equation:
\begin{equation} \label{eq:hamilton_jacobi}
    \frac{\partial \phi}{\partial \lambda} + \frac{1}{2} \|\nabla \phi\|^2 = \frac{\psi}{p} \, .
\end{equation}
Equations \eqref{eq:potential_flow} and \eqref{eq:hamilton_jacobi}, combined with the continuity equation \eqref{eq:continuity}, form the complete system describing the optimal Bayesian transport.

\begin{table*}[tb]
\centering
\renewcommand{\arraystretch}{1.5}
\setlength{\tabcolsep}{8pt}
\caption{{Comparison of Variational Particle Flow, Standard QM, and Bohmian Mechanics. Structural and conceptual comparison, separating the abstract variational particle flow from the two distinct physical interpretations of the identical quantum hydrodynamic equations.} \\ 
}
\label{tab:comparison}

\begin{tabular}{@{}p{2.8cm} p{3.5cm} p{3.5cm} p{3.5cm}@{}}
\toprule
\textbf{Feature} & \textbf{Variational Particle Flow \newline (Bayesian Inference)} & \textbf{Standard QM \newline (Madelung Hydrodynamics)} & \textbf{Bohmian Mechanics \newline (Pilot Wave Theory)} \\
\midrule

\textbf{System} &  Abstract probability transport & Abstract probability evolution & Real physical particles guided by the pilot wave \\

\textbf{Evolution parameter} & Pseudo-time ($\lambda \in [0, 1]$) & Physical time ($t$) & Physical time ($t$) \\

\textbf{Primary entity} & Probability density $p(x, \lambda)$ (Posterior belief) & Wave Function $\Psi(x,t)$ (Probability amplitude) & Particle Position $X(t)$ \& Pilot Wave $\Psi(x,t)$ \\

\textbf{Probability density} &  Posterior probability & Probability of finding the particle &  Actual distribution of real particles \\


\textbf{Velocity field} & {Transport velocity ($v=\nabla \phi$):} \newline Maps prior to posterior & {Probability flow ($v=\nabla S/m$):} \newline Rate of flow of probability mass (statistical) & {Particle velocity ($v=\nabla S/m$):} \newline The velocity of particles in space \\

\textbf{Driving potential}  & {Information potential} ($-\frac{\psi}{p}$): \newline Enforces Bayes' Rule. & Localization / Uncertainty kinetic energy & Quantum potential force field \newline A non-local force guiding the particle \\

\textbf{Reality (Ontology)} & {Computational:} \newline Particles are just samples & {Indeterminate:} \newline Particles have no position until measured & {Deterministic:} \newline Particles always have precise trajectories \\

\textbf{Role of ``Interference''} & N/A & Probability amplitudes cancelling out & The $Q$ potential steering particles into bands \\

\bottomrule
\end{tabular}
\end{table*}

\section{Geometric and physical interpretation} \label{sec:interpret}

The variational framework derived in Section~\ref{sec:var} establishes a mathematical structure for the optimal particle flow. In this section, we analyze the geometric properties of this flow and its isomorphism to quantum hydrodynamics, which provides a dynamical perspective on the Bayesian update.

\subsection{Geometric duality: Fisher--Rao and Wasserstein}
The proposed framework reveals a connection between statistical inference and optimal transport. The log-homotopy constraint defines a one-dimensional exponential family connecting the prior and posterior densities. In information geometry \cite{amari2016information}, this path constitutes an $e$-geodesic on the statistical manifold endowed with the Fisher--Rao metric.

While the density evolution is fixed by the $e$-geodesic, the velocity field moving the probability mass is not. The continuity equation therefore defines an equivalence class of admissible velocity fields that generate the same marginal density evolution. Within this class, the variational principle selects a distinguished irrotational representative by minimizing the kinetic action subject to the homotopy constraint. By minimizing the kinetic energy action $S$, we invoke the dynamic formulation of optimal transport \cite{benamou2000computational}. The resulting potential flow $v = \nabla \phi$ is the Wasserstein-$2$ optimal transport map. Thus, in our framework the Wasserstein optimal transport flow is constrained to track a {Fisher-Rao geodesic}.

\subsection{Similarity to Madelung hydrodynamics}
The coupled system of the continuity equation \eqref{eq:continuity} and the Hamilton--Jacobi equation \eqref{eq:hamilton_jacobi} reveals a mathematical isomorphism to the hydrodynamic formulation of quantum mechanics, originally introduced by Madelung \cite{madelung1927quantentheorie} and later central to the de Broglie--Bohm interpretation \cite{bohm1952suggested}.

To see this, consider the time-dependent Schrödinger equation for a single particle of mass $m$ in an external potential $V(x)$:
\begin{equation} \label{eq:schrodinger}
    i\hbar \frac{\partial \Psi}{\partial t} = \left( -\frac{\hbar^2}{2m}\nabla^2 + V(x) \right) \Psi \, ,
\end{equation}
where $\hbar=h/2\pi$ is the reduced Planck constant and $i$ is the imaginary unit.
Madelung proposed the ansatz where the complex wavefunction $\Psi$ is decomposed into the polar form
\begin{equation}
    \Psi(x,t) = \sqrt{\rho(x,t)} \exp\left( \frac{i}{\hbar} S(x,t) \right) \, ,
\end{equation}
where $\rho(x,t) = |\Psi|^2$ represents the probability density, and $S(x,t)$ is the action (or phase) with units of angular momentum.

Substituting this ansatz into \eqref{eq:schrodinger} and separating the real and imaginary components yields two coupled non-linear partial differential equations describing the evolution of the quantum fluid:

\begin{enumerate}
    \item {Imaginary part (continuity equation):}
    \begin{equation}
        \frac{\partial \rho}{\partial t} + \nabla \cdot \left( \rho \frac{\nabla S}{m} \right) = 0 \, .
    \end{equation}
    This ensures the conservation of probability mass, identifying the flow velocity as $v = \nabla S / m$.
    
    \item {Real part (quantum Hamilton--Jacobi equation):}
    \begin{equation} \label{eq:quantum_HJ}
        \frac{\partial S}{\partial t} + \frac{1}{2m}\|\nabla S\|^2 + V + Q = 0 \, .
    \end{equation}
\end{enumerate}

Equation \eqref{eq:quantum_HJ} differs from the classical Hamilton--Jacobi equation solely by the addition of the term $Q$, known as the quantum potential:
\begin{equation}
    Q(x,t) = -\frac{\hbar^2}{2m} \frac{\nabla^2 \sqrt{\rho}}{\sqrt{\rho}} \, .
\end{equation}
Unlike the external classical potential $V(x)$, the quantum potential $Q$ is intrinsic as it depends on the curvature of the probability density itself. In Bohmian mechanics, $Q$ generates a non-local force that guides particles, preventing trajectories from crossing and giving rise to interference phenomena (e.g., the double-slit experiment).

\begin{figure*}[ht]
    \centering
    \includegraphics[width=\textwidth]{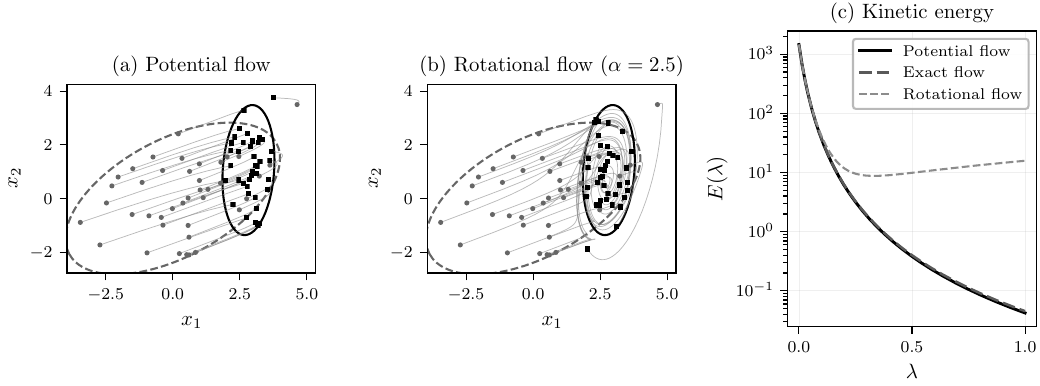}
    \caption{Numerical validation on a 2D linear Gaussian Bayesian update with a correlated prior $P_0$. (a)~Particle trajectories under the optimal potential flow: paths are smooth and direct. (b)~Trajectories under the rotational flow ($\alpha = 2.5$): paths are curved yet all particles arrive at the same posterior (solid ellipse), confirming that both flows produce the exact Bayesian update. (c)~Instantaneous kinetic energy $E(\lambda)$ on a logarithmic scale. The potential flow (solid) achieves the lowest energy at every $\lambda$, followed closely by the log-homotopy exact flow (long-dashed), with the rotational flow incurring the highest kinetic cost.}
    \label{fig:trajectories}
\end{figure*}

\subsubsection{The Bayesian analogy}
We now map this quantum framework to our variational particle flow. By setting the physical constants $\hbar=1$ and $m=1$, and mapping physical time $t$ to pseudo-time $\lambda$, our derived system becomes structurally identical to the Madelung equations.

Comparing \eqref{eq:hamilton_jacobi} with \eqref{eq:quantum_HJ}, we identify the correspondence:
\begin{equation}
    -\frac{\psi(x,\lambda)}{p(x,\lambda)} \iff V(x) + Q(x,t) \, .
\end{equation}
The term $\psi/p$ acts as the total effective potential driving the Bayesian update. This provides a physical interpretation: just as the quantum potential $Q$ exerts a force to shape the probability density, the Lagrange multiplier term $\psi$ acts as an information potential. It constrains the otherwise unconstrained transport to follow the log-homotopy density path prescribed by Bayes' theorem.

Table \ref{tab:comparison} lists analogies between the log-homotopy potential flow and quantum mechanics.

\subsection{Dynamical analysis: the Euler equation}
We can shift from a kinematic description ($v=\nabla \phi$) to a dynamical one by deriving the acceleration field. Applying the gradient operator $\nabla$ to the Hamilton-Jacobi equation \eqref{eq:hamilton_jacobi}:
\begin{equation} \label{eq:grad_HJ}
    \nabla \left( \frac{\partial \phi}{\partial \lambda} \right) + \nabla \left( \frac{1}{2} \|v\|^2 \right) = \nabla \left( \frac{\psi}{p} \right) \, .
\end{equation}
Using Schwarz's theorem $\nabla \frac{\partial \phi}{\partial \lambda} = \frac{\partial v}{\partial \lambda}$ and the vector identity $\nabla (\frac{1}{2} \|v\|^2) = (v \cdot \nabla)v+ v \times (\nabla \times v)$, we obtain the material derivative of the velocity:
\begin{equation} \label{eq:euler_equation}
    \frac{Dv}{D\lambda} \triangleq \frac{\partial v}{\partial \lambda} + (v \cdot \nabla)v = \nabla \left( \frac{\psi}{p} \right) \, .
\end{equation}
Equation \eqref{eq:euler_equation} is the Euler equation for an inviscid fluid. The term $\nabla(\psi/p)$ represents the force field generated by the information potential. This provides a mechanism for particle acceleration: if the density  deviates from the log-homotopy path, $\psi$ increases, creating a gradient that forces particles back to the optimal trajectory.

\subsection{Computational implications}
The dynamical formulation offers advantages over purely kinematic approaches for high-dimensional filtering.

\subsubsection{Symplectic integration}
Standard particle flow filters use first-order Euler integration ($x_{k+1} = x_k + \epsilon v_k$), which introduces discretization errors. By computing the acceleration $a = \nabla(\psi/p)$ via \eqref{eq:euler_equation}, we can apply second-order symplectic integrators (e.g., velocity Verlet). These methods use the momentum of the flow to take larger pseudo-time steps $\Delta \lambda$ while improving energy conservation and trajectory accuracy.

\subsubsection{Stiffness detection}
The magnitude of the information force $\|\nabla(\psi/p)\|$ serves as a metric for the numerical stiffness of the update. Large forces indicate a significant mismatch between the current particle distribution and the target homotopy. It can be used to indicate the need for adaptive step-size reduction. Conversely, small forces allow for accelerated integration steps.

\subsubsection{Physics-Informed Regularization}
In learning-based implementations (e.g., PINNs), the Hamilton-Jacobi residual acts as a physical regularizer. Minimizing the residual
\begin{equation}
    \mathcal{R}_{HJ} = \left\| \frac{\partial \phi}{\partial \lambda} + \frac{1}{2}\|\nabla \phi\|^2 - \frac{\psi}{p} \right\|^2
\end{equation}
ensures the learned map is not merely a geometric interpolation but a valid conservative flow, preventing the network from learning solutions that violate probability conservation.

\section{Numerical experiment} \label{sec:exp}

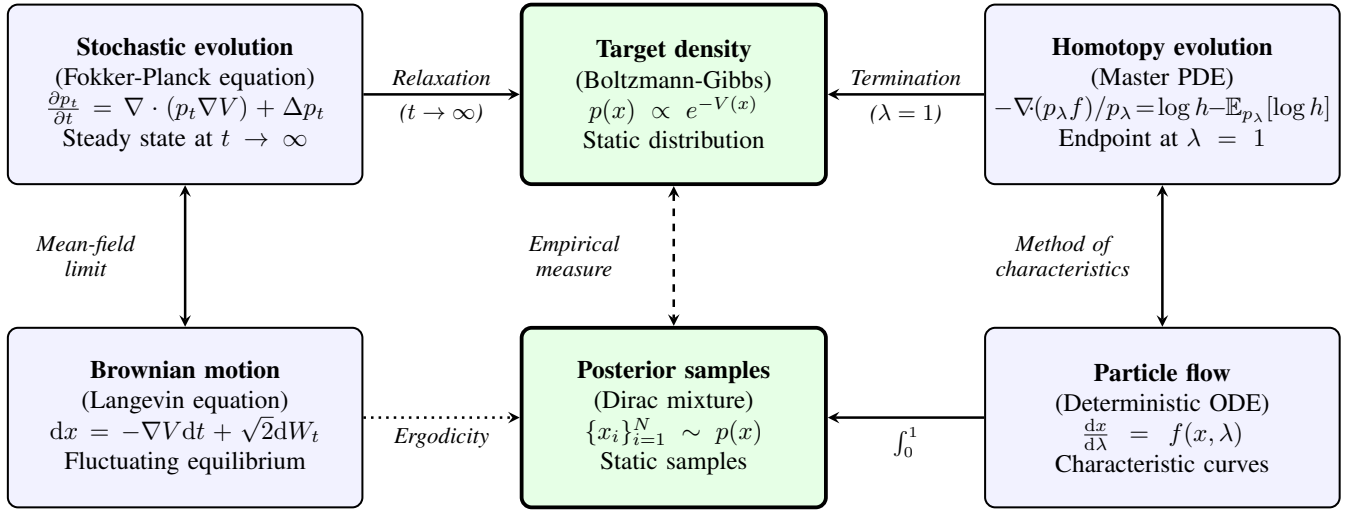
\begin{figure*}[ht]
    \centering
    \resizebox{\textwidth}{!}{
    \begin{tikzpicture}[node distance=6.8cm, auto, thick]
        \tikzstyle{block} = [rectangle, draw, fill=blue!5, 
                             text width=4.7cm, text centered, rounded corners, minimum height=2.5cm]
        \tikzstyle{target} = [rectangle, draw, fill=green!10, 
                             text width=4cm, text centered, rounded corners, minimum height=2.5cm, line width=1.5pt]
        \tikzstyle{arrow} = [->, >=stealth, line width=1pt]
        
        \tikzstyle{txt} = [text width=2.5cm, align=center, font=\small\itshape]
        \tikzstyle{lbl} = [align=center, font=\small\itshape] 

        \node [target] (target_pdf) {\textbf{Target density}\\(Boltzmann-Gibbs)\\$p(x) \propto e^{-V(x)}$\\Static distribution};
        
        \node [block, left of=target_pdf] (stoch_pdf) {\textbf{Stochastic evolution}\\(Fokker-Planck equation)\\$ \frac{\partial p_t}{\partial t} = \nabla \cdot (p_t \nabla V) + \Delta p_t$\\Steady state at $t \to \infty$};
        
        \node [block, right of=target_pdf] (homo_pdf) {\textbf{Homotopy evolution}\\(Master PDE)\\ $- \nabla \!\!\cdot \!(p_\lambda f)/p_\lambda \!=\! \log h \!-\! \mathbb{E}_{p_\lambda}[\log h]$ \\ Endpoint at $\lambda = 1$};

        \node [target, below of=target_pdf, node distance=4.5cm] (target_part) {\textbf{Posterior samples}\\(Dirac mixture)\\$\{x_i\}_{i=1}^N \sim p(x)$\\Static samples};

        \node [block, below of=stoch_pdf, node distance=4.5cm] (stoch_part) {\textbf{Brownian motion}\\(Langevin equation)\\$\diff x = -\nabla V \diff t + \sqrt{2} \diff W_t$\\Fluctuating equilibrium};

        \node [block, below of=homo_pdf, node distance=4.5cm] (homo_part) {\textbf{Particle flow}\\(Deterministic ODE)\\$\frac{\diff x}{\diff \lambda} = f(x,\lambda)$\\Characteristic curves};

        \draw [arrow, <->] (stoch_pdf) -- node[txt, left] {Mean-field\\limit} (stoch_part);
        
        \draw [arrow, <->, dashed] (target_pdf) -- node[txt, left] {Empirical\\measure} (target_part);
        
        \draw [arrow, <->] (homo_pdf) -- node[txt, left] {Method of\\characteristics} (homo_part);

        \draw [arrow] (stoch_pdf) -- node[lbl, above] {Relaxation} node[lbl, below] {($t \to \infty$)} (target_pdf);
        
        \draw [arrow, dotted] (stoch_part) -- node[txt, sloped, below] {Ergodicity} (target_part);

        \draw [arrow] (homo_pdf) -- node[lbl, above] {Termination} node[lbl, below] {($\lambda = 1$)} (target_pdf);

        \draw [arrow] (homo_part) -- node[txt, sloped, below] {$\int_0^1$} (target_part);

    \end{tikzpicture}
    }
    \caption{Physics-based Bayesian computation. Center: The estimation objective is a static target posterior and its discrete approximation via an empirical measure. Left: Stochastic relaxation, where the system asymptotically converges to a fluctuating equilibrium ($t \to \infty$). The macroscopic and microscopic views are linked via the mean-field limit. Right: Deterministic transport, where the system evolves over a finite horizon ($\lambda \in [0,1]$). The particle trajectories correspond to the characteristic curves of the macroscopic master PDE.}
    \label{fig:inference_landscape}
\end{figure*}

To validate the theoretical claims, we present a numerical study on a 2D linear Gaussian Bayesian update that admits exact, closed-form solutions. We construct three velocity fields that each satisfy \eqref{eq:master_eqn} and therefore produce the identical posterior density, but carry different kinetic energy costs.

\subsection{Setup}
We consider a 2D state $x \in \mathbb{R}^2$ with  Gaussian prior $g(x) = \mathcal{N}(0, P_0)$, where
\begin{equation}
    P_0 = \begin{bmatrix} 4 & 1.5 \\ 1.5 & 2 \end{bmatrix} .
\end{equation}
The measurement model is $z = Hx + \varepsilon$, $\varepsilon \sim \mathcal{N}(0, R)$, with $H = [1,\; 0]$, $R = 0.25$, and observed value $z = 3$.
Under the log-homotopy \eqref{eq:homotopy}, the intermediate density remains Gaussian, $p(x,\lambda) = \mathcal{N}(\mu(\lambda), P(\lambda))$, with
\begin{align}
    P(\lambda)^{-1} &= P_0^{-1} + \lambda H^\top R^{-1} H, \\
    \mu(\lambda) &= P(\lambda)\bigl(P_0^{-1}\mu_0 + \lambda H^\top R^{-1} z\bigr).
\end{align}
This yields the analytic posterior $\mu(1) = [2.82,\; 1.06]^\top$ and $P(1) = \operatorname{diag}(0.24,\; 1.47)$.

\subsection{Flows under comparison}

For the Gaussian homotopy density, we seek an affine velocity field of the form $v = M(\lambda)(x - \mu(\lambda)) + \frac{\diff\mu}{\diff\lambda}$.
Substituting this ansatz into the continuity equation \eqref{eq:continuity} and matching coefficients by polynomial degree in $(x-\mu)$ yields a single constraint on the Jacobian matrix $M(\lambda)$:
\begin{equation} \label{eq:lyapunov_constraint}
    P(\lambda)^{-1} M + M^\top P(\lambda)^{-1} = -H^\top R^{-1} H \, .
\end{equation}
Equation \eqref{eq:lyapunov_constraint} is underdetermined: any matrix $M$ that satisfies it generates a valid flow. The three flows below are particular solutions of \eqref{eq:lyapunov_constraint} with different $M$.

\subsubsection{Potential flow (variational optimum)}
An affine flow $v = M(x-\mu) + c$ is irrotational if and only if its spatial Jacobian $\partial v / \partial x = M$ is symmetric, since $(\nabla \times v)_{ij} = M_{ij} - M_{ji}$. The symmetric solution of \eqref{eq:lyapunov_constraint} is obtained from the continuous Lyapunov equation:
\begin{equation} \label{eq:lyapunov}
    P(\lambda)^{-1} S + S\, P(\lambda)^{-1} = -H^\top R^{-1} H \, .
\end{equation}
Since $P(\lambda)^{-1}$ is positive definite, the Lyapunov operator $\mathcal{L}(X) = P^{-1}X + XP^{-1}$ is invertible on symmetric matrices, guaranteeing a unique symmetric solution $S(\lambda)$.

The resulting velocity $v_\mathrm{pot} = S(\lambda)(x - \mu(\lambda)) + \frac{\diff\mu}{\diff\lambda}$ satisfies $\nabla \times v_\mathrm{pot} = 0$ and is therefore the irrotational potential flow predicted by \eqref{eq:potential_flow}.

\subsubsection{Exact log-homotopy flow (EF)}
The closed-form solution \cite{daum2010exact,ding2012implementation} yields $v_\mathrm{EF} = A(\lambda)x + b(\lambda)$, with
\begin{align}
    A(\lambda) &= -\tfrac{1}{2} P_0 H^\top \!\left(\lambda H P_0 H^\top + R\right)^{-1}\! H \, , \label{eq:DH_A}\\
    b(\lambda) &= \left(I + 2\lambda A\right)\!\left[\left(I + \lambda A\right) P_0 H^\top R^{-1} z + A\mu_0\right] \label{eq:DH_b} .
\end{align}
By the matrix inversion lemma, $A(\lambda) = -\frac{1}{2}P(\lambda)H^\top R^{-1}H$, and one can verify that $A$ satisfies the constraint \eqref{eq:lyapunov_constraint}. However, for $H = [1, 0]$, the matrix $A$ takes the structure $A = \sigma(\lambda)\bigl[\begin{smallmatrix}P_{11} & 0\\ P_{21} & 0\end{smallmatrix}\bigr]$ for a scalar $\sigma(\lambda)$, which is symmetric only when $P_{21}=0$, i.e., when the prior is uncorrelated. For the correlated prior considered here, $\|A - A^\top\|_F = 0.47$ at $\lambda = 0.5$, confirming that the exact flow contains a residual rotational component and is therefore suboptimal in the kinetic energy sense.

\subsubsection{Explicitly rotational flow}
A solenoidal perturbation can always be added to any valid flow without altering the density evolution. We use:
\begin{equation}
    v_\alpha = v_\mathrm{pot} + \alpha\, J \nabla \log p(x, \lambda) \, ,
\end{equation}
where $J = \bigl[\begin{smallmatrix}0 & 1\\-1 & 0\end{smallmatrix}\bigr]$ is the $90^\circ$ rotation matrix and $\alpha \in \mathbb{R}$ is the rotation strength.
Since $\nabla \cdot (J \nabla p) = -\partial^2 p / \partial x_1 \partial x_2 + \partial^2 p / \partial x_2 \partial x_1 = 0$ by Schwarz's theorem, the added term contributes nothing to the divergence $\nabla \cdot (p v_\alpha)$, so $v_\alpha$ satisfies \eqref{eq:master_eqn} for any $\alpha$.
By construction the total action decomposes exactly as $\mathcal{S}(\alpha) = \mathcal{S}^* + c\alpha^2$ with $c = \frac{1}{2}\int_0^1\!\operatorname{tr}(P(\lambda)^{-1})\diff\lambda$, guaranteeing a global minimum at $\alpha = 0$.

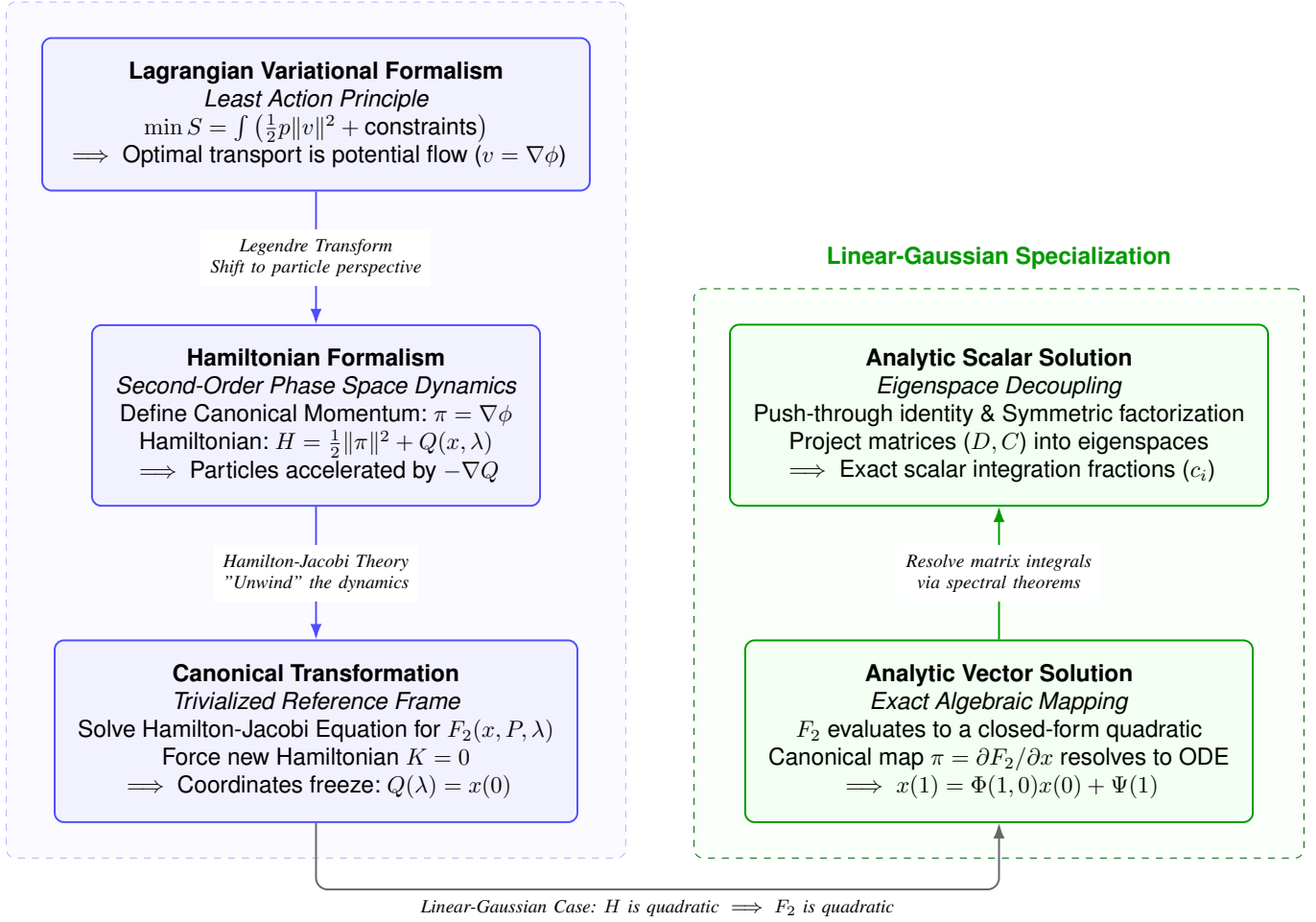
\begin{figure*}[t]
\centering
\resizebox{\textwidth}{!}{
\begin{tikzpicture}[
    >=Latex,
    node distance=1.5cm and 2.5cm, 
    font=\sffamily,
    theoryBox/.style={
        rectangle, rounded corners, draw=blue!70, fill=blue!5, 
        thick, minimum width=6cm, minimum height=1.8cm, align=center,
        inner sep=10pt
    },
    exactBox/.style={
        rectangle, rounded corners, draw=green!60!black, fill=green!5, 
        thick, minimum width=6cm, minimum height=1.8cm, align=center,
        inner sep=10pt
    },
    labelNode/.style={
        fill=white, font=\footnotesize\itshape, align=center, inner sep=4pt
    },
    groupLabel/.style={
        font=\bfseries\sffamily, text=#1
    }
]


\node[theoryBox] (lagrange) {
    \textbf{Lagrangian Variational Formalism}\\
    \textit{Least Action Principle}\\
    $\min S = \int \left(\frac{1}{2}p\|v\|^2 + \text{constraints}\right)$\\
    $\implies$ Optimal transport is potential flow ($v = \nabla \phi$)
};

\node[theoryBox, below=2cm of lagrange] (hamilton) {
    \textbf{Hamiltonian Formalism}\\
    \textit{Second-Order Phase Space Dynamics}\\
    Define Canonical Momentum: $\pi = \nabla \phi$\\
    Hamiltonian: $H = \frac{1}{2}\|\pi\|^2 + Q(x,\lambda)$\\
    $\implies$ Particles accelerated by $-\nabla Q$
};

\node[theoryBox, below=2cm of hamilton] (canonical) {
    \textbf{Canonical Transformation}\\
    \textit{Trivialized Reference Frame}\\
    Solve Hamilton-Jacobi Equation for $F_2(x,P,\lambda)$\\
    Force new Hamiltonian $K=0$\\
    $\implies$ Coordinates freeze: $Q(\lambda) = x(0)$
};


\node[exactBox, right=2.5cm of canonical] (vector) {
    \textbf{Analytic Vector Solution}\\
    \textit{Exact Algebraic Mapping}\\
    $F_2$ evaluates to a closed-form quadratic\\
    Canonical map $\pi = \partial F_2 / \partial x$ resolves to ODE\\
    $\implies x(1) = \Phi(1,0) x(0) + \Psi(1)$
};

\node[exactBox, above=2cm of vector] (scalar) {
    \textbf{Analytic Scalar Solution}\\
    \textit{Eigenspace Decoupling}\\
    Push-through identity \& Symmetric factorization\\
    Project matrices ($D, C$) into eigenspaces\\
    $\implies$ Exact scalar integration fractions ($c_i$)
};


\draw[->, thick, draw=blue!70] (lagrange) -- node[labelNode] {Legendre Transform\\ Shift to particle perspective} (hamilton);
\draw[->, thick, draw=blue!70] (hamilton) -- node[labelNode] {Hamilton-Jacobi Theory\\ "Unwind" the dynamics} (canonical);

\draw[->, thick, draw=green!60!black] (vector) -- node[labelNode] {Resolve matrix integrals\\ via spectral theorems} (scalar);


\draw[->, thick, draw=black!60, rounded corners=5pt] 
    (canonical.south) -- ++(0,-1cm) -| node[labelNode, pos=0.25, yshift=-8pt] {Linear-Gaussian Case: $H$ is quadratic $\implies F_2$ is quadratic} (vector.south);


\begin{scope}[on background layer]
    \node[rectangle, draw=blue!30, fill=blue!2, dashed, rounded corners, 
          fit=(lagrange) (hamilton) (canonical), inner sep=15pt] (theoryBg) {};
    \node[above=5pt of theoryBg.north, groupLabel=blue!80!black] {General Theory (Valid for Nonlinear Systems)};

    \node[rectangle, draw=green!40!black, fill=green!2, dashed, rounded corners, 
          fit=(vector) (scalar), inner sep=15pt] (exactBg) {};
    \node[above=5pt of exactBg.north, groupLabel=green!60!black] {Linear-Gaussian Specialization};
\end{scope}

\end{tikzpicture}
}
\caption{From Lagrangian variational formalism to Hamiltonian mechanics}
\label{fig:canonical}
\end{figure*}

\subsection{Results}

Fig.~\ref{fig:trajectories} shows 40 simulated particle trajectories and the instantaneous kinetic energy $E(\lambda) = \int \frac{1}{2} p \|v\|^2 \diff x$.
For the linear affine flow $v = M(x-\mu) + \frac{\diff\mu}{\diff\lambda}$, this evaluates analytically as $E(\lambda) = \frac{1}{2}\bigl[\operatorname{tr}(M^\top\! M P(\lambda)) + \|\frac{\diff\mu}{\diff\lambda}\|^2\bigr]$, enabling exact energy comparisons without Monte Carlo approximation.
All three flows deliver every particle to the correct posterior, confirming that any solution of \eqref{eq:lyapunov_constraint} produces the exact Bayesian update. However, the rotational flow incurs substantially higher kinetic energy at every $\lambda$, visible in panel~(c).

The resulting total actions are $\mathcal{S}^* = 31.72$ (potential), $\mathcal{S}_\mathrm{EF} = 31.92$ (exact flow), and $\mathcal{S}_{2.5} = 41.23$ (rotational, $\alpha=2.5$). The numerical results are consistent with the variational minimum-action characterization of the potential flow. The exact flow has a non-symmetric Jacobian $A$: the antisymmetric part of $A$ contributes excess kinetic energy that the variational principle eliminates.

\section{Conclusion and future directions} \label{sec:conc}

In this work, we applied a variational Lagrangian framework to address the non-uniqueness of the exact particle flow filter. By modeling the probability density as a pressureless inviscid fluid and applying the principle of least action, we obtained an optimal transport map that minimizes the kinetic energy of the transformation. This derivation yielded several theoretical insights: the optimal velocity field is necessarily irrotational, the flow corresponds to a Wasserstein-2 optimal transport constrained to a Fisher-Rao geodesic, and the resulting Hamilton-Jacobi equations are formally analogous to Madelung's hydrodynamic formulation of quantum mechanics. Furthermore, deriving the material acceleration shifted the problem from a kinematic to a dynamical description, providing a basis for stiffness analysis and higher-order symplectic integration.

Building upon this established Hamiltonian formalism, our proposed framework opens the door to advanced analytical techniques adapted from classical mechanics for future research. As illustrated in Fig.~\ref{fig:inference_landscape}, while stochastic evolution relies on asymptotic relaxation, the log-homotopy particle flow constructs a deterministic ordinary differential equation over a finite pseudo-time horizon. Future extensions can explore canonical transformations (Fig.~\ref{fig:canonical}) by solving the Hamilton-Jacobi equation to construct a type-2 generating function, $F_2(x, P, \lambda)$. This transforms the phase space into a trivialized reference frame where the new Hamiltonian $K$ equals zero, effectively freezing the dynamics such that $Q(\lambda) = x(0)$. The entire particle trajectory can then be recovered by inverting the canonical mapping.

This theoretical direction is particularly promising when specialized to the linear-Gaussian case. Because the original Hamiltonian $H$ takes a quadratic form, the corresponding generating function $F_2$ is inherently quadratic. This guarantees that the canonical momentum mapping $\pi = \partial F_2 / \partial x$ resolves into an exact algebraic mapping, allowing deterministic flow integration without discretization errors. By projecting the system matrices into decoupled eigenspaces using symmetric factorization, exact scalar integration fractions ($c_i$) can be extracted directly via spectral theorems. Future work will leverage these exact algebraic mappings to develop novel control variates and exact flow regularizers for highly nonlinear, high-dimensional target distributions.

Alongside these canonical extensions, future research will also explore generalizing the Lagrangian action to Riemannian manifolds for state estimation on constrained geometries. Practical implementations will investigate mesh-free methods, such as reproducing kernel Hilbert spaces, to solve the coupled Hamilton–Jacobi system without grid-based PDE solvers. Finally, formal links between the log-homotopy variational flow and the stochastic control formulations used in diffusion-based generative modeling (such as the Schrödinger bridge problem) will be investigated.

\bibliographystyle{ieeetr}
\bibliography{ref}

\vspace{12pt}

\end{document}